# Using Machine Learning in Analyzing Air Quality Discrepancies of Environmental Impact

Paul Wang[*†] Lucas Yang[†], Radhouane Chouchane[*], Jin Guo[*], Michael A. Bailey[‡]
[*]Morgan State University
{paul.wang, radhouane.chouchane, jin.guo}@morgan.edu
[†]University of Maryland
{paulwang, lucasyang}@umd.edu
[‡]Georgetown University
{michael.bailey}@georgetown.edu

*Abstract*—In this study, we apply machine learning and software engineering in analyzing air pollution levels in City of Baltimore. The data model was fed with three primary data sources: 1) a biased method of estimating insurance risk used by homeowners' loan corporation, 2) demographics of Baltimore residents, and 3) census data estimate of NO2 and PM2.5 concentrations. The dataset covers 650,643 Baltimore residents in 44.7 million residents in 202 major cities in US. The results show that air pollution levels have a clear association with the biased insurance estimating method. Great disparities present in NO2 level between more desirable and low income blocks. Similar disparities exist in air pollution level between residents' ethnicity. As Baltimore population consists of a greater proportion of people of color, the finding reveals how decades old policies has continued to discriminate and affect quality of life of Baltimore citizens today.

## I. Introduction

A growing body of scholarship has conducted research showing that there exists a relationship between racially discriminatory legal practices of the past and disparities in the quality of life for United States citizens today [1]–[4]; in particular, that mortgage appraisal practices of past decades are influencing the quality of air for residents of varying races and ethnicities [5], [6].

Regarding the aforementioned appraisal practices, following the onset of the Great Depression, the Home Owners' Loan Corporation (henceforth referred to as HOLC) was created in 1933 to assist borrowers in fulfilling mortgage payments and financing property purchases [7]. In order to achieve this, part of the HOLC's activities included assigning neighborhoods in the country a letter grade on a four-point scale, with "A" meaning "Best," "B" meaning "Still Desirable," "C" meaning "Definitely Declining," and "D" meaning "Hazardous" [8] (residents living in areas with poorer grades were less likely to be approved for loans); however, what comes under scrutiny is the criteria under which these grades were assigned, with some explicitly referencing the race/ethnicity of residents as cause to give lower grades (redlining) [9].

This research is funded by a grant from the Bezos Earth Fund (#438424).

Previously, a study has been conducted in which nationwide HOLC grade data was combined with statistics on $NO_2$ and $PM_{2.5}$ levels, along with demographic information pulled from the 2010 United States Census, to create plots displaying the association between HOLC grade and air pollution levels, and the association between race/ethnicity and air pollution levels [5]. This paper contains plots generated using the same sources of data, with a focus on the Baltimore, Maryland location, and will corroborate previous studies' findings that historical discriminatory practices continue to influence the contemporary state of the environment that civilians live in [10], [11]. Applying machine learning in cyber threat analysis was conducted by the authors previously [12], [13].

## II. Support Vector Machine and Random Forests Models

Decision tree is one of the common models to extracting classification from featured instances. The issue with generalization makes it unfavorable to Support Vector Machine (SVM) in machine learning [14].

### A. SVM Model

Let $S = \{(x_1, y_1), (x_2, y_2), ..., (x_N, y_n)\}$ be a training dataset, where $x_i \in R^N$ and $y_i \in \{-1, 1\}$ for $i = 1, 2, ..., N$. The hyper-plane of S is defines as

$$f(x) = (\sum_{j=1}^{N} y_j a_j^0 x_j \cdot x) + b_0$$

$$= (\sum_{i=1}^{N} y_i a_i^0 (x_i x) + b_0$$

The optimal is reached when $f(x) = 0$.

The model is trained using k-means clustering. For each data partition, computing using SVM and find the margin until an optimal margin is reached [15]–[20].

### B. Random Forests Algorithm

Random Decision Forests improve prediction with better generalization outcomes. The algorithm randomly divides training data into subsets each generate a random tree to construct a random forest.

## III. METHODOLOGY AND KNOWLEDGE DISCOVERY

### A. Data Cleaning and Pre-processing

Data need to go through the cleaning and pre-processing before feeding into data models. This step includes: reducdant information, strong correlations, temporal pattern, and outlier analysis.

The initial step in data transformation involved comprehensive data cleaning to address various issues. This process was aimed at enhancing the quality of the data, making it more suitable for analytical models. The following tasks were performed during this phase:

- Structural Transformation: Unnecessary columns, including those with a single value or irrelevant to the study's objectives, such as unit parameter name and code, and method name, were removed. The columns were also renamed to ensure clarity and consistency throughout the dataset.
- Temporal Filtering: All records preceding the last date saved in the existing tables were removed to avoid data redundancy.
- Geographical Filtering: Since this study only concerns the USA, data entries pertaining to locations outside the United States, specifically Canada, Puerto Rico, and Mexico, were identified and excluded from the dataset.
- Data Integrity Correction: Negative concentration values, resulting from flawed calibration, were removed.
- Data Type Standardization: State and county codes were converted into integers, and dates were transformed into the datetime.
- Data Separation: The concentration data was separated from AQI data for easier visualization in the next steps.

### B. Programming for Data Processing

We wrote scripts in Python3 to process raw data sources before filtering them to produce plots displaying the association between HOLC grade, race/ethnicity residents, and air pollution levels in Baltimore.

Data from the 2010 United States Census was assembled in a Pandas DataFrame containing information from 202 major American cities amounting to $n_{national}$ = 2,023,728 blocks, with 44,776,346 residents living in HOLC-mapped areas. This was then filtered to only include entries pertaining to blocks within Baltimore, Maryland ($n_{local}$ = 10,036 blocks, with 650,643 residents in HOLC-mapped areas).

### C. Air Pollution Data Analysis

After preparing the data for processing according to the above-mentioned methodology, air pollution levels were collected and presented in three formats:

1) Unadjusted: Air pollution levels in blocks in Baltimore by HOLC grade and race/ethnicity.
2) Intraurban Adjusted: The difference between air pollution levels experienced by residents in blocks in Baltimore and the overall population-weighted mean air pollution level in Baltimore. This emphasizes the air quality in any given block in Baltimore relative to local conditions.
3) Nationally Adjusted: The difference between air pollution levels experienced by residents in blocks in Baltimore and the national population-weighted mean air pollution level. This emphasizes the air quality in any given block in Baltimore relative to the national state of affairs.

$NO_2$ levels are measured in ppb, and $PM_{2.5}$ levels in micrograms per cubic meter ($\mu g/m^3$).

## IV. RESULTS AND ANALYSIS

### A. HOLC Grades and Concentration in Baltimore

Discussion of unadjusted and nationally adjusted $NO_2$ levels are shown in Figures 1 and 2, and unadjusted and nationally adjusted $PM_{2.5}$ levels in Figures 3 and 4. Bars represent 25th and 75th percentiles. Medians are indicated with horizontal lines, means with dots in each bar, and the overall mean is shown by the dotted line.

The left cluster displays $NO_2$ levels by HOLC grade, and the right cluster displays $NO_2$ levels by race/ethnicity.

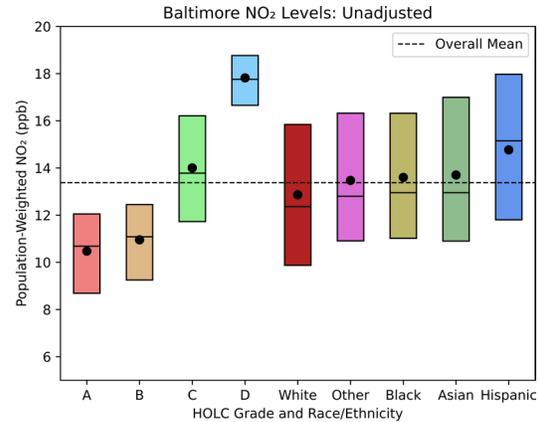

Fig. 1. Unadjusted population-weighted distribution of $NO_2$ levels in HOLC-mapped blocks in Baltimore.

Figure 1 displays unadjusted air pollution levels in Baltimore, showing that redlining is strongly associated with $NO_2$ concentration, especially in the more "hazardous" grades, C, and D. $NO_2$ levels in the 'C' grade, with an interquartile range of 11.73 ppb to 16.21 ppb, lies almost completely above the ranges of both grades 'A' and 'B'. $NO_2$ levels in the 'D' grade are even higher, with an interquartile range of 16.66 ppb to 18.77 ppb, which, despite being a smaller range, not only exceeds that of any other HOLC grade, but also any other air pollution measurement by race/ethnicity. As is evident, $NO_2$ levels by race/ethnicity also follow an upwards trend, but is less pronounced than the HOLC grouping, with median $NO_2$ levels for White, Other, Black, and Asian residents being nearly consistent (12.36 ppb, 12.80 ppb, 12.95 ppb, and 12.95 ppb, respectively).

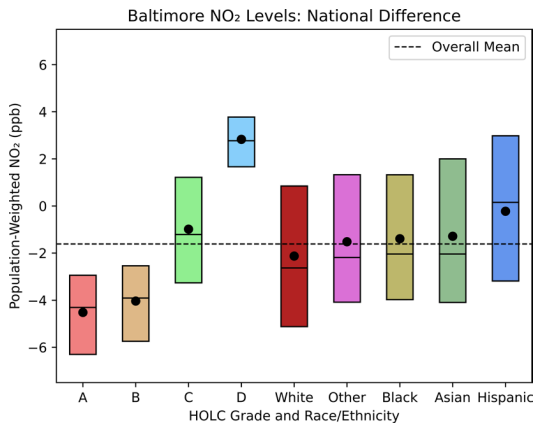

Fig. 2. Differences in $NO_2$ levels between Baltimore and national population-weighted mean.

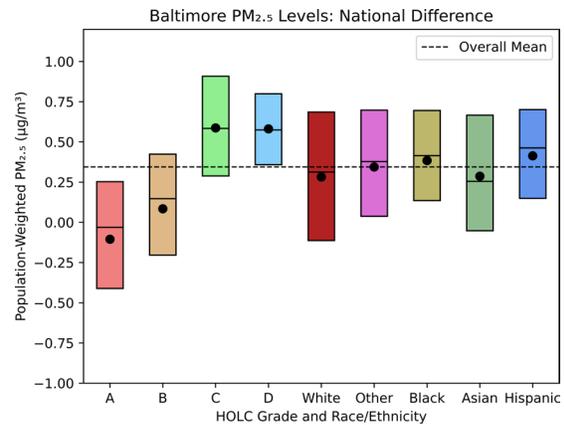

Fig. 4. Differences in $PM_{2.5}$ levels between Baltimore and national population-weighted mean.

Figure 2 displays a similar pattern in $NO_2$ measurements in Baltimore, but presents them as the difference between air pollution levels in Baltimore from the national population-weighted mean level. Corroborating the results from Figure 1, $NO_2$ levels follow HOLC grades more closely, with the median levels in the 'A' and 'B' grades being 4.31 ppb and 3.91 ppb lower than the national mean, and in the 'D' grade, 2.77 ppb higher.

In addition, Figure 4 shows that the distributions of $PM_{2.5}$ levels across race/ethnicity all exhibit similar 75th percentile measurements (all fall between $11.30 \pm 0.03$ $\mu g/m^3$), but show greater differences in 25th percentile values; that is, the distribution of air pollution levels experienced by White and Asian residents indicate greater range and lower floor measurements than those of Black or Hispanic residents.

Figure 4 supports these findings by showing that the distribution of $PM_{2.5}$ levels in 'A' and 'B' graded neighborhoods measure lower than the national mean, whereas 'C' and 'D' neighborhoods measure greater than the national mean, and $PM_{2.5}$ levels by race/ethnicity do not display as monotonic a trend.

### B. Disparities by Race/Ethnicity in Baltimore

Figures 5 and 6 display intraurban differences in $NO_2$ and $PM_{2.5}$ levels in Baltimore by both HOLC grade and ethnicity; each line represents air pollution measurements for some ethnicity in the data, subdivided into four points, one for each HOLC grade.

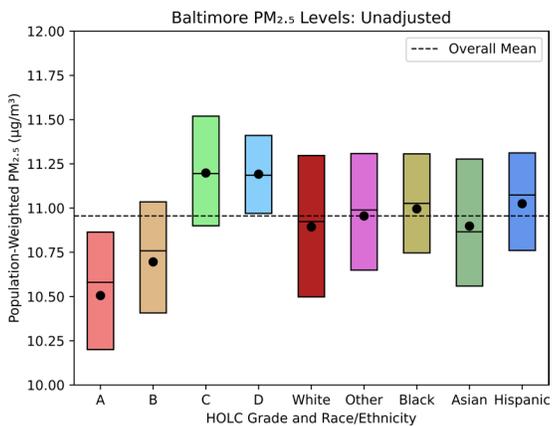

Fig. 3. Unadjusted population-weighted distribution of $PM_{2.5}$ levels in HOLC-mapped blocks in Baltimore.

Figure 3 displays unadjusted $PM_{2.5}$ levels in Baltimore. Though this plot demonstrates that $PM_{2.5}$ levels are greater in areas with poorer HOLC grades (neighborhoods with grades 'A' and 'B' both fall nearly squarely below the overall mean of 10.96 $\mu g/m^3$), $PM_{2.5}$ levels in neighborhoods with a 'C' grade exhibit both a larger interquartile range than those in the 'D' grade (0.62 $\mu g/m^3$ against 0.44 $\mu g/m^3$), and a greater 75th percentile measurement (11.52 $\mu g/m^3$ to 11.41 $\mu g/m^3$).

Figure 5 shows that in Baltimore, $NO_2$ levels are greater at worse HOLC grades for all ethnicities, with White and Asian residents generally experiencing lower levels than Black or Hispanic residents. In neighborhoods with an 'A' grade, there is a greater difference in $NO_2$ levels among residents of varying ethnicities, with those of color bearing the brunt of it; Black residents living in 'A' graded neighborhoods experience an $NO_2$ concentration that is 1.72 ppb below the population-weighted mean concentration of Baltimore, while White residents in 'A' neighborhoods live in areas with $NO_2$ levels 4.04 ppb below that average. However, at the lowest HOLC grade, 'D,' the differences in $NO_2$ levels among the varying races/ethnicities of residents decrease, with measurements converging at $4.50 \pm 0.40$ ppb above the mean Baltimore measurement.

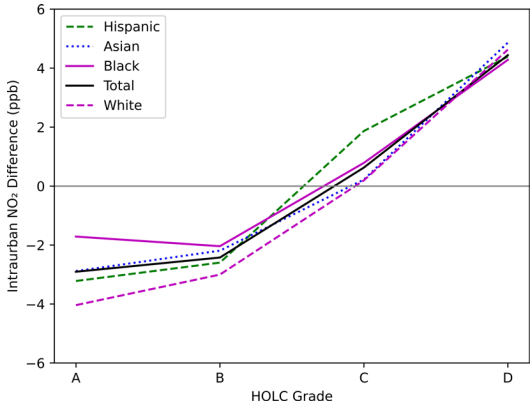

Fig. 5. Distribution of intraurban differences in $NO_2$ levels in HOLC-mapped blocks by ethnicity in Baltimore.

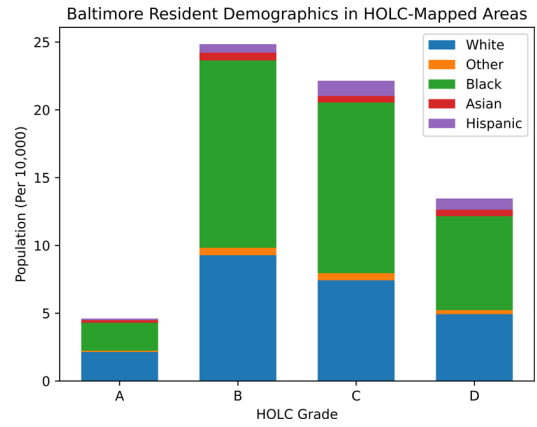

Fig. 7. Number of residents living in neighborhoods by HOLC grade in Baltimore.

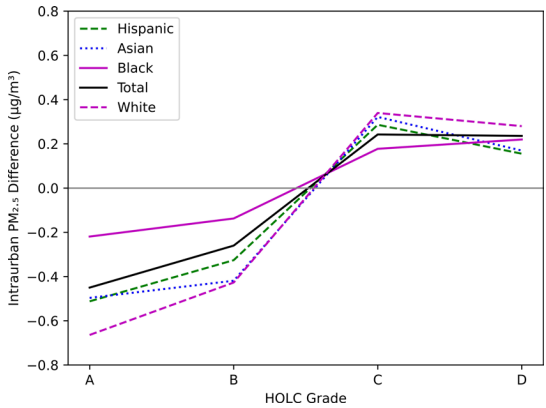

Fig. 6. Distribution of intraurban differences in $PM_{2.5}$ levels in HOLC-mapped blocks by ethnicity in Baltimore.

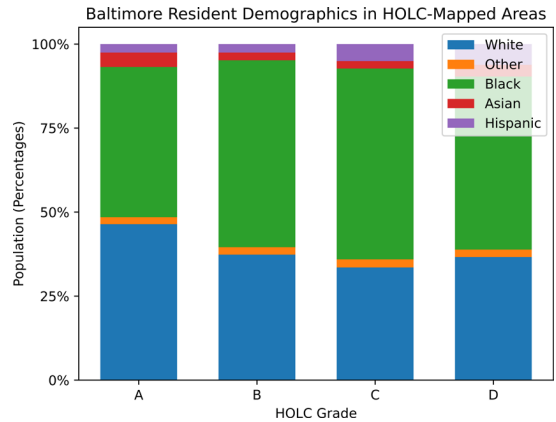

Fig. 8. Proportion of residents living in HOLC-mapped neighborhoods in Baltimore of races/ethnicities.

Figure 6 shows that in Baltimore, $PM_{2.5}$ levels are, in general, greater at worse HOLC grades for ethnicities, with the following specification: at "better" HOLC grades, especially 'A,' there is a more significant difference in $PM_{2.5}$ measurements among ethnicities, with residents of color experiencing higher $PM_{2.5}$ levels than White or Asian residents. This trend is reversed at the 'C' grade, where $PM_{2.5}$ measurements for Black residents are 0.18 $\mu g/m^3$ above the Baltimore mean, while $PM_{2.5}$ measurements for White residents are 0.34 $\mu g/m^3$ above the Baltimore mean.

### C. Baltimore Demographics

Figures 7, 8, and 9 display demographic statistics for residents living in Baltimore, Maryland.

Figure 7 displays the number of residents, expressed as multiples of 10,000, living in Baltimore, and how many live in neighborhoods of each HOLC grade. The majority ($\geq 70\%$) of Baltimore residents live in neighborhoods assigned a 'B' or 'C' grade, with fewer in 'D' neighborhoods, and only 7% in 'A' neighborhoods.

Figure 8 expounds on Figure 7 by presenting the proportion of the population in each HOLC-mapped neighborhood by ethnicity; according to the data, the majority of residents living in Baltimore are White or Black, with the proportion of residents of color increasing among poorer HOLC grades.

Figure 9 compares the distribution of demographic proportions in Baltimore with nationwide statistics. Across all HOLC grades, there are a greater number of Black residents than people of any other ethnicity, with differences ranging from 20% to 35% more Black residents in Baltimore than the nationwide mean. Figure 9 also reveals that the proportion of White residents in neighborhoods in Baltimore compared to nationwide increase as HOLC grade worsens, while the proportion of Hispanic residents decreases.

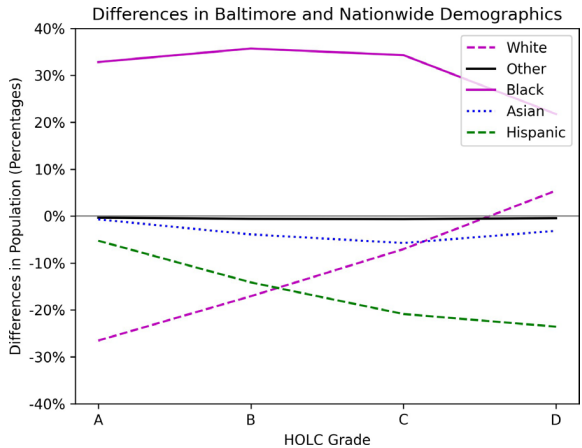

Fig. 9. Differences in proportions of residents of ethnicities between Baltimore and United States.

## V. Dashboard Design and Development

Starting from city of Baltimore, we are in the process of expanding the study scope to the State of Maryland and Nationwide. A Dashboard is a very effective tool for Human Computer Interactions (HCI). The design and development process encompasses a series of steps aimed at creating an efficient and user-friendly platform for visualizing and analyzing air quality data. It consists of initial design, backend infrastructure, UI development, and testing procedures.

### A. Interactive Maps

A Levels of Details of interactive map was developed using the Syncfusion library. This map (Figure 10) allows for various interaction to explore detailed air quality data including: a) zoom, focus on State, view details of a State, County, and City, air quality evolution over time, air quality by season, years, months, days, and hours.

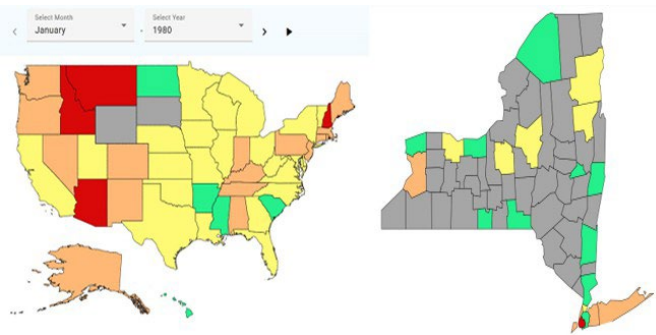

Fig. 10. An interactive map on the dashboard front page shows the air pollution data of the United State and a State when clicking on the US map.

### B. Backend Architecture

The backend architecture was selected prioritizing simplicity and a clear separation of concerns to streamline development and ensure smooth data retrieval and endpoint exposure. Given the straightforward nature of the backend operations, which primarily involve fetching data from the database and exposing endpoints to serve aggregated data and statistics, Flask was chosen for the development of the endpoints.

Flask is known for its simplicity and flexibility, making it an ideal choice for projects with relatively uncomplicated backend requirements. With Flask, it was possible to quickly set up the backend infrastructure and define endpoints without unnecessary configurations.

Furthermore, Flask's lightweight nature and minimalistic approach allowed us to keep our backend codebase concise and focused. This simplicity not only enhances development speed but also facilitates easier debugging and maintenance in the long run.

### C. API Development

API development for frontend-backend communication and data retrieval begins with the identification of endpoints required to serve the frontend application's needs. In the provided code, this involves defining routes in app.py that correspond to specific data queries or operations, such as retrieving average pollution values (*/average value*), fetching row counts (*/count*), or obtaining air quality category information (*/air quality category*). These routes are designed to accept parameters from frontend requests, such as the pollutant element, year, month, state, and county, using the request object in Flask.

Once the routes are defined, the next step is to implement the corresponding controller logic, where service functions from service.py are invoked to interact with the database models and perform data retrieval or processing tasks. The service functions encapsulate the business logic and database operations, abstracting away the complexities of querying the database and performing calculations.

Error handling is implemented to handle cases where invalid data or no data is found for a given query, ensuring robustness and reliability in the API responses. Finally, the API responses are formatted as JSON objects and returned to the frontend application, enabling seamless communication between the frontend and backend components of the application.

### D. Project Deployment

The system is deployed to AWS using the DevSecOps methodology. The DevSecOps process starts with code commits to the GitHub repositories *aqi-quality-apis* for the backend and *aqi-dashboard* for the frontend. Each commit triggers AWS CodePipeline, which orchestrates the entire workflow, providing a streamlined architecture.

For the backend, AWS CodePipeline uses AWS CodeBuild to compile the Flask application, run tests, and create deployment artifacts. Successful builds are deployed to AWS Elastic Beanstalk, which simplifies the management of the application environment by handling provisioning, scaling, and monitoring. AWS Elastic Beanstalk integrates seamlessly with AWS RDS, ensuring efficient database management.

For the frontend, AWS CodePipeline also uses AWS CodeBuild to compile the Angular application, run tests, and produce build artifacts. These artifacts are then deployed to Amazon S3, where they are hosted as a static website. Amazon S3 offers a reliable and scalable solution for static web hosting, ensuring that the frontend is quickly and efficiently delivered to users.

This automated pipeline ensures consistent integration, security [21]–[23] and deployment of both backend and frontend components, minimizing manual intervention and reducing the risk of errors. The workflow is illustrated in the figure 11, providing a visual overview of the deployment pipeline and demonstrating the efficiency of the AWS-based architecture.

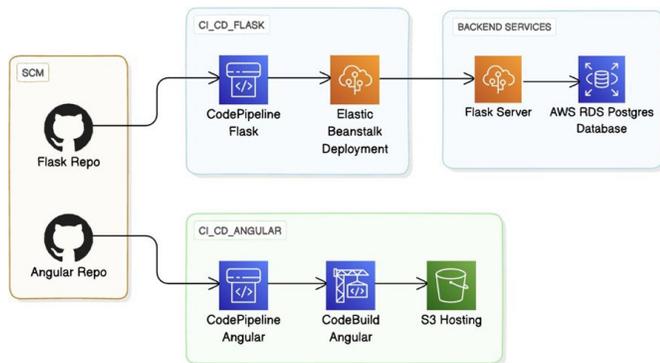

Fig. 11. Application deployment architecture and services on AWS

## VI. Implications and Future Directions

Results from our analysis yields key insight into air pollution levels and their relationship with racially charged policy-making of the past in Baltimore, Maryland. First, that HOLC rankings of decades past hold a clear association with differences in present-day environmental quality. Second, that HOLC grades influence air pollution levels more heavily than race/ethnicity, and that this influence is more clearly defined in $NO_2$ levels than in $PM_{2.5}$ levels. However, this does not discount the link between ethnicity and air pollution levels, and the finding that Baltimore has a greater proportion of residents of color than nationwide, as well as observable differences in $NO_2$ levels, corroborates previous bodies of research on the subject [5]. Since HOLC grades of the past continue to influence inequities today, local propositions to address these topics may take into account the unevenly impacted communities and structural changes that may have further effects over many more years.